\documentclass{aastex62}

\graphicspath{{./}{figures/}}

\shorttitle{Maximum Drift Rates}
\shortauthors{Sheikh et al.}

\begin{document}

\title{Choosing a Maximum Drift Rate in a SETI Search: Astrophysical Considerations}

\author[0000-0001-7057-4999]{Sofia Z. Sheikh}
\address{Department of Astronomy \& Astrophysics and \\ Center for Exoplanets and Habitable Worlds\\ 525 Davey
Laboratory, The Pennsylvania State University, University Park, PA, 16802, USA}

\author[0000-0001-6160-5888]{Jason T.\ Wright}
\address{Department of Astronomy \& Astrophysics and \\ Center for Exoplanets and Habitable Worlds\\ 525 Davey
Laboratory, The Pennsylvania State University, University Park, PA, 16802, USA}

\author[0000-0003-2828-7720]{Andrew Siemion}
\address{Department of Astronomy, University of California, Berkeley, 501 Campbell Hall 3411, Berkeley, CA, 94720, USA}
\affiliation{Department of Astrophysics/IMAPP, Radboud University, P.O. Box 9010, NL-6500 GL Nijmegen, The Netherlands}
\affiliation{SETI Institute, Mountain View, CA 94043, USA}

\author[0000-0003-2516-3546]{J. Emilio Enriquez}
\address{Department of Astronomy, University of California, Berkeley, 501 Campbell Hall 3411, Berkeley, CA, 94720, USA}
\affiliation{Department of Astrophysics/IMAPP, Radboud University, P.O. Box 9010, NL-6500 GL Nijmegen, The Netherlands}

\begin{abstract}

A radio transmitter which is accelerating with a non-zero radial component with respect to a receiver will produce a signal that appears to change its frequency over time. This effect, commonly produced in astrophysical situations where orbital and rotational motions are ubiquitous, is called a drift rate. In radio SETI (Search for Extraterrestrial Intelligence) research, it is unknown a priori which frequency a signal is being sent at, or even if there will be any drift rate at all besides motions within the solar system. Therefore a range of potential drift rates need to be individually searched, and a maximum drift rate needs to be chosen. The middle of this range is zero, indicating no acceleration, but the absolute value for the limits remains unconstrained. A balance must be struck between computational time and the possibility of excluding a signal from an ETI. In this work, we examine physical considerations that constrain a maximum drift rate and highlight the importance of this problem in any narrowband SETI search. We determine that a normalized drift rate of 200 nHz (eg. 200 Hz/s at 1 GHz) is a generous, physically motivated guideline for the maximum drift rate that should be applied to future narrowband SETI projects if computational capabilities permit.

\end{abstract}


\section{Introduction} \label{sec:intro}

An electromagnetic transmitter which is approaching or receding from a receiver at a constant velocity will produce a signal that will be measured at a blueshifted or redshifted frequency relative to the rest frequency of the transmitter. A transmitter that is accelerating radially with respect to the receiver will thus produce a signal whose frequency changes over time. This ``drift rate'', or ``chirp'', commonly measured in Hz/s, is a necessary parameter for any radio Search for Extraterrestrial Intelligence (SETI) observation. If the transmitter is in an exoplanetary system, the characteristics of the star and exoplanets in the system could affect the resulting drift rate.

Since the early 1960's, the radio and microwave frequency bands have been suggested as the ideal place to look for intentional beacons\footnote{The benefits of broadcasting at radio wavelengths apply equally well to unintentional signals} from extraterrestrial intelligence due to information transfer at the speed of light, low energy-per-bit costs, and low attenuations by matter in the galaxy \citep{cocconi1959searching}. Given that the radio and microwave bands present numerous frequencies to search, many suggestions have been put forth for Shelling points \citep{schelling1960strategy} in frequency space, often referred to as ``magic frequencies''; these are wavelengths that have some additional reason for an ETI to use, such as the 21cm line of neutral hydrogen or the broader water hole \citep{oliver1979rationale}. In these early SETI projects only a limited range of frequencies could be searched at one time due to the capacities of the receivers. With modern wideband receivers, the exact choice of frequency is less pressing. Unfortunately, even if a perfect frequency or frequency band can be determined, it is unknown what signal modulation would indicate communication from an ETI.
	
The simplest possibility is to look for a signal with very high signal-to-noise ratio (SNR) with a very small frequency width; signals produced by nature have a minimum width due to natural and thermal broadening \citep{cordes1997scintillation}. Masers from sources such as star-forming regions are the narrowest persistent astrophysical features that have been detected in the radio wavelengths, with widths on the order of hundreds of Hz. Human radio technology, even in the 1960's, could easily produce signals with widths on the order of Hz \citep{tarter2001search}. Under this assumption, creating instrumentation with extremely narrow frequency channels, covering an extremely large bandwidth \citep[as used in Breakthrough Listen, see][]{macmahon2017recorder}, gives the searcher the best chance of detecting a narrowband signal \citep{papagiannis1985recent}.

However, in an extra complication, a Hz/s drift rate will cause the power from the signal to span multiple frequency channels over the course of the observation. The narrower the channels are and the greater the drift rate of the signal, the less power in each individual channel (decreasing the signal to noise) and the more distorted the original signal shape if the drift is unaccounted for.

The solution to this problem is to ``un-smear'' the data - to correct for the drift rate in an attempt to maximize the SNR. This process must be repeated for many drift rates and the results compared because it is not known a priori what drift rate the signal will have. But how many drift rates should be searched, and bounded by what maximum drift rate? These are the questions that this paper seeks to answer.

Previous observational searches for SETI signals with the Green Bank Telescope \citep{siemion20131, enriquez2017breakthrough, margot2018search} and the Allen Telescope Array (ATA) \citep{harp2016seti, tarter2011first} have defined a maximum drift rate somewhat arbitrarily and then occasionally determined the properties of the exoplanet that such a drift rate would accommodate. In addition, SETI assets such as the MCSA \citep{cullers1985iau} and SETI@Home \citep{anderson2002seti} have necessitated work on how to computationally handle the search for drift rates, but without special consideration of their physical basis. An exception to this is the review of \citet{oliver1971project} which used the drift rate of an Earth-like planet with an 8 hour day, thereby accidentally defining a common literature standard used by some of the searches above. \citet{sullivan1978eavesdropping} considered the Doppler drift of signals leaving the Earth, while \citet{cornet2003solar} pointed out that correlating ``anomalous microwave phenomena'' observed with telescopes such as the ATA to known drift rates within the solar system could allow for localization of a signal to a particular body. \citet{harp2016seti} notes that orbital or rotational motion of a transmitter in an exoplanetary system can produce a drift rate, but does not quantify what upper limits on this motion might be. In this paper we will take a different approach from these previous studies by calculating maximum drift rates from physically allowable accelerations, informed by knowledge from the last decade of exoplanetary discoveries.

In Section \ref{sec: math}, we give a didactic presentation of the theory of drift rates and link them to astrophysical sources. We discuss computational considerations for searches in Section \ref{sec: computation} in an elementary style to provide context for the later discussion. We compute different maximum drift rates in Section \ref{sec: mdrexamples} and discuss the benefits and consequences of these maxima in Section  \ref{sec: discussion}. Finally, we summarize and conclude in Section \ref{sec: conclusion}.

\section{Theory of Drift Rates} \label{sec: math}

\subsection{Drift Rates \label{ssec: drift}}


In this section, we will discuss the physics and mathematics necessary for understanding drift rates. For simplicity, we will assume that a constant-frequency, narrowband signal is being sent, as observed in the reference frame of the  transmitter; however, the underlying physical causes of drift rates will apply similarly to any signal modulation. Recall that any drift that is observed in the signal on Earth is due to a change in relative velocities between the transmitter and the observer, or, said another way, the difference in reference frames between the transmitter and the observer. Our goal is to identify the transformation that will yield a minimal deviation from the strength and shape of the signal in the reference frame of the transmitter. Performed incorrectly, this could cause a real signal to be left unflagged in the data. The timescale here is relevant: the longer the observation is performed, the smaller the drift rate that will start to have observable effects on the data if uncompensated for.

\subsubsection{Determining the Relevant Terms}
\label{sssec: relevant_terms}

In its most general form, a drift rate can be expressed as the time derivative of a Doppler Shift: 

\begin{equation}
\label{eq: doppler}
\dot{f} = \frac{dv_r}{dt} \frac{f_{rest}}{c}
\end{equation}

\noindent Where \(\dot{f}\) is the drift rate, \(f_{rest}\) is the rest frequency of the signal, and \(\frac{dv_r}{dt}\) is the total relative radial acceleration between the transmitter and receiver. 

The total relative radial acceleration can be expressed generally as a sum of various radial components. 

\begin{equation}
    \label{eq: sumform}
    \frac{dv_r}{dt} = \Sigma_i \frac{dv_{r,i}}{dt}
\end{equation}

The terms summed on the right side will be dependent on the exact configuration of the transmitting system - a free-floating transmitter will have different terms than a transmitter on a host body, which will have different terms from a self-propelled transmitter. This sum could therefore include:

\begin{enumerate}
    \item Rotational acceleration from the Earth
    \item Orbital acceleration of the Earth about the Sun (including effects from eccentricity)
    \item Rotational acceleration from a body on which a transmitter is placed
    \item Orbital acceleration of a body on which a transmitter is placed or orbital acceleration of a free-floating transmitter (including effects from eccentricity)
    \item Additional acceleration variation on 1 from the Earth's oblateness and topology or on 3 from the host body's oblateness and topology
    \item Acceleration from galactic potential
    \item Rotational acceleration from a transmitter itself
    \item Acceleration from a transmitter with propulsion
    \item Gravitational redshift in the solar system
\end{enumerate}

Cosmological accelerations are extremely small, scaling as $czH_0$, so the terms above will also apply to intergalactic signals.

Actually calculating this quantity requires a physical understanding of the system and a sense of which terms we can neglect. In this work, we are looking at order-of-magnitude effects and can therefore ignore many of these terms which will not affect the final answer. 

The drift rate imparted solely by a receiver placed on the surface of the Earth is dominated by the rotational acceleration of the Earth for most lines of sight: the orbital motion's contribution is less than 2\% of the rotational motion's contribution at the equator when observing the horizon. When generalizing to other systems, however, the orbital contribution could be significant (see Section \ref{sec: mdrexamples}). To investigate the effect of Earth's eccentricity, we can compute the difference in drift rate at Earth's periapse and apoapse with Equation \ref{eq: general_orbit}; using orbital parameters from \citet{Simon1994} the maximum difference from the circular case is 6.7\%. For completeness, we must account for the Earth's orbital contribution in a non-circular way. Likewise, for other systems, the eccentricity component cannot be assumed to be negligible. Therefore terms 1-4 are essential.

Acceleration from Earth's oblateness and topology can be neglected given our order-of-magnitude approach. Oblateness and topology of the host body will be neglected in our formulation both for the same reason and because there are no data on these characteristics for exoplanetary systems. Hence, term 5 will not be considered.

Since galactic acceleration is orders of magnitude smaller than rotational acceleration, relative motions between star systems are negligible and we do not include them. To motivate this, the length of the galactic year is 233 million years \citep{innanen1978interaction}, and the distance from the Sun to the center of the galaxy is 8.0 kiloparsecs \citep{reid1993distance}. Plugging into Equation \ref{eq: general_orbit}, we find that the galactic acceleration from the sun's motion around the center of the galaxy is approximately \(1.8 \times 10^{-10}\) m/s\(^2\). The acceleration from Earth's rotation is approximately 0.03 m/s\(^2\), or eight orders of magnitude larger. We will thus ignore term 6.

Acceleration from a transmitter with propulsion, though constrainable by the speed of light, cannot be informed by measurable astronomical parameters - we will include term 7 in the equation in Section \ref{sssec: all_term_eq} but will not attempt to evaluate it further.

The effect of gravitational redshift in the solar system on the drift rate is negligible and the rotational motion of a transmitter itself is unconstrainable; we consider neither in this work, eliminating terms 8 and 9.

Given the terms that survive, it can be seen that planetary parameters can be related to the maximum drift rate. 

\subsubsection{Rotational Contribution}
\label{sssec: rotationalcont}

Acceleration from circular motion can be described by the following equation:

\begin{equation}
\label{eq: circularmotion}
\frac{dv}{dt}_{circ} = \frac{v_{eq}^2}{R} \mathrm{sin}(\theta)\mathrm{sin}(i)
\end{equation}

Here, \(v_{eq}\) is the equatorial velocity of the rotating object, $R$ is the radius of the rotating object (assuming a sphere), \(\theta\) is the co-latitude in the object's coordinates, and $i$ is the sky-plane inclination of the system.

In order to maximize the radial acceleration, and therefore the drift rate, \(\mathrm{sin}(\theta)\) and \(\mathrm{sin}(i)\) will be set to unity, modeling a transmitter on the equator of the hypothetical planet and an inclination of 90\(^{\circ}\). Thus \(\frac{dv}{dt}_{max} = \frac{v_{eq}^2}{R}\), which gives us the maximum drift rate from one component of rotational acceleration. We convert the equatorial velocity to period and radius using the circumference \(v_{eq} = \frac{2 \pi R}{P}\) to obtain:

\begin{equation}
\label{eq: circularmotion_max}
\frac{dv}{dt}_{circ, max} = \frac{4 \pi^2 R}{P_{rot}^2} 
\end{equation}

\subsubsection{Orbital Contribution}
\label{sssec: orbital_cont}

Acceleration under gravity for an orbiting object is simply given by:

\begin{equation}
    \label{eq: general_orbit}
    \frac{dv}{dt}_{grav, max} = \frac{G M_{central}}{r^2}
\end{equation}

\noindent $M_{central}$ is the mass of the central object and $r$ is the distance to the center of that object. The equation gives the maximum drift rate: from the observing angle where all of the acceleration is radial. This orbital formulation applies to both transmitters on host bodies and free-floating transmitters and is independent of eccentricity and period. 

\subsubsection{Equation of All Dominant Terms}
\label{sssec: all_term_eq}

Given the explanations above, we are left with four dominant components to the drift rate: the rotation of Earth, the orbital motion of Earth, the rotation of the host body and the orbital motion of the host body. We include an ``other'' term to account for acceleration from a transmitter with propulsion (term 8 in Section \ref{sssec: relevant_terms}).

\begin{equation}
\label{eq: master}
\dot{f} = \frac{f_{rest}}{c} \left(\frac{4 \pi^2 R_{\oplus}}{P_{rot, \oplus}^2} + \frac{4 \pi^2 R}{P_{rot}^2} + \frac{G M_{sun}}{{r_{\oplus}}^2} + \frac{G M_{central}}{r^2} + \frac{dv}{dt}_{other}\right)
\end{equation}

\noindent Here r and r$_{\oplus}$ are the orbital distances of the transmitter's host body (or transmitter) and the Earth, R and R$_{\oplus}$ are the radii of the transmitter's host body and the Earth, and P$_{rot}$ and P$_{rot, \oplus}$ are the rotation period of the transmitter's host body and the Earth.

For a sense of scale, the maximum drift rate at $f_{rest} = 8$ GHz for the Green Bank Telescope (top of the C-band receiver \citep{gbt2017proposer}) is 0.91 Hz/s due to the Earth's rotational and orbital motion. If both sides are divided by $f_{rest}$, the drift rate becomes independent of transmission frequency; later in this work we will report this normalized quantity in the unit of nanoHertz (nHz).

\subsection{Break-Up Speeds \label{ssec: breakup}}

The break-up rotation rate, or \textit{break-up speed}, is the rate of rotation at which the centrifugal force at the equator balances the gravitational force at the equator, disrupting the structural integrity of a gravitationally bound body. Given that the break-up speed is the largest rotation rate a planet can have, this grants a first-order upper limit for rotational rates. One way to calculate the period at which the break-up occurs is to set the rotational velocity at the equator equal to the velocity that the surface material would have if it were actually orbiting at the radius of the object. This limit will not apply to small bodies (such as those in Section \ref{sssec: NEANEO}) that are held together by electrostatic forces instead of gravitational ones. For a spherically symmetric, uniformly dense, gravitationally bound body subject to no other forces and without differential rotation, the orbital velocity at the equator is given by:

\begin{equation}
\label{eq: orbital}
v_{orb} = \sqrt{\frac{GM}{R}}
\end{equation}

Where M is the mass of the object and R is the radius of the object. Rotational velocity on the equator is given by:

\begin{equation}
\label{eq: rotational}
v_{rot} = \frac{2 \pi R}{P_{rot}}
\end{equation}

Where $P_{rot}$ is the rotation period. Combining these equations, we get an expression for the break-up period:

\begin{equation}
\label{eq: Pbreakup}
P_{breakup} = \frac{2 \pi R^{3/2}}{\sqrt[]{GM}} = \sqrt{\frac{3\pi}{G\rho}}
\end{equation}

For the Earth, the associated rotation period is about 84.5 minutes. More complicated models that address the assumption of sphericity including deformation and other effects can be found in e.g. \citet{bertotti1990rotation}. 

In Equation \ref{eq: Pbreakup}, $\rho$ refers to the bulk density of the object. The break-up speed depends only on mass enclosed (bulk density), not the density distribution. Real planets, however, have nonuniform density distributions from differentiation and compression from gravitational forces \citep{Seager2007}. When considering the density distribution of an exoplanet using something like the Adams-Williamson equation \citep{Williamson1923, Dziewonski1981}, the mass scales much more sharply with radius than a uniform density distribution would estimate. More massive planets are therefore generally denser, implying that the most massive planet of a given composition will have the shortest break-up period. To give a sense of the scale of this effect, the radius of a 6$M_\oplus$ planet made entirely of iron is a factor of 1.5 larger than it should be if calculated with an assumption of constant density (using a nonuniform density estimate from \citet{Seager2007}).

Current core-accretion and terrestrial planet formation models suggest that we should expect near break-up speed rotational velocities to be common \citep{kokubo2007formation, batygin2018terminal}. However, this is not seen in any known system, even for objects as large as brown dwarfs \citep{bryan2018constraints, scholz2018universal}. This paradox has been investigated before \citep[e.g., ][]{batygin2018terminal} but a consensus has yet to be reached.

\section{Computational and Algorithmic Considerations \label{sec: computation}}

One solution to the problem of choosing a maximum drift rate is to try every physically possible drift rate, even those corresponding to relativistic objects, and avoid the problem of choosing a maximum entirely.\footnote{This is the spirit of the approach taken by SETI@home (which searches the maximum drift rate permitted by the data dimensions) \citep{anderson2002seti}.} Although this is tempting because it is ideal to minimize the number of assumptions, searching through all possible drift rates, even on a grid, quickly becomes very computationally expensive.\footnote{In fact, there is a second additional constraint: allowing extremely high drift rates leaves your algorithm vulnerable to RFI from LEO satellites, which can easily hit hundreds of Hz/s at 1.5 GHz \citep{harp2016seti}. However, other forms of RFI rejection should be able to eliminate these signals.}

\begin{figure}[ht]
\centering
\includegraphics[width=0.6\textwidth]{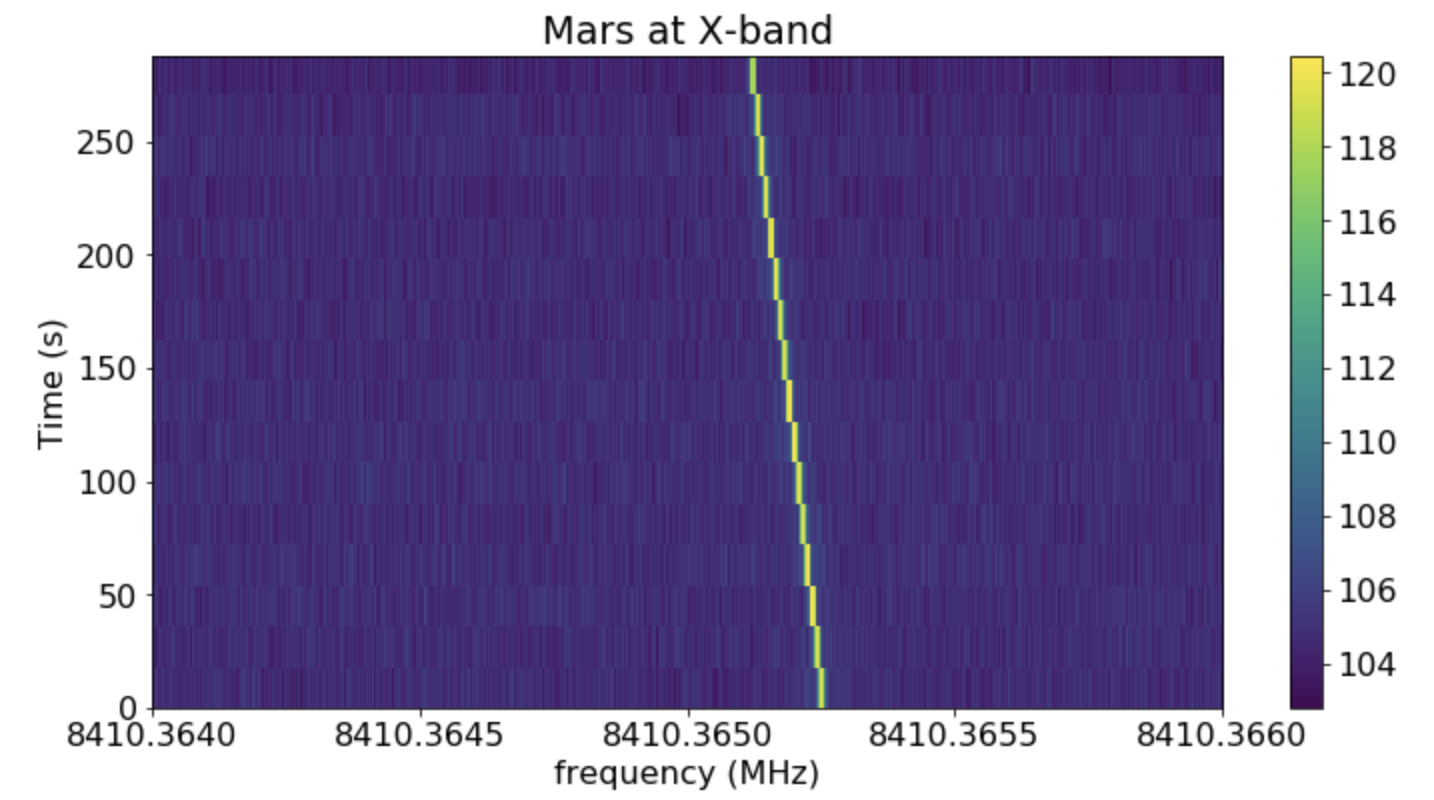}
\caption{A waterfall plot of a real artificial narrowband signal from human technology at Mars, observed with the GBT 8.0-11.6 GHz (X-band) receiver and the Breakthrough Listen back-end \citep{macmahon2017recorder}. The y-axis shows observation time and the x-axis shows the frequency in MHz. The signal drifts linearly towards lower frequencies over the duration of the observation. The color of each elongated pixel shows the intensity in dB, as shown in the color bar to the right.} 
\label{img: mars_improved}
\end{figure}

The data that must be searched is in the form of a ``waterfall plot'', also called a dynamic spectrum. These waterfall plots can be represented in a single image displaying a received power for each frequency at each time. An example waterfall plot containing a drifting signal from extraterrestrial human technology is shown in Figure \ref{img: mars_improved}. The full range of frequencies in the data is the bandwidth of the observation, and the bandwidth is sampled at some frequency resolution to produce a discrete frequency axis. The full time range is the observation duration, and the duration is broken into smaller integrations at some time resolution to produce a discrete time axis. As with any time-frequency dataset, there is a classical uncertainty trade-off between achievable time resolution and achievable frequency resolution; for narrowband SETI applications, extremely high frequency resolution (to discriminate between artificial signals and those from natural maser emission) is preferred at the expense of time resolution. 

The Breakthrough Listen (BL) project provides a quantitative example of the data structure. The BL data pipeline begins with time series of raw voltages recorded at the telescope. Because of the storage constraints and to facilitate further analysis, the raw voltage data are turned into three dynamic spectra of varying time and frequency resolution. The highest frequency resolution data, for SETI applications, are produced with a 2.7 Hz frequency resolution and an 18 second time resolution \citep{macmahon2017recorder}. A 5 minute observation with this data product produces an image of approximately $16 \times 10^9$ pixels per GHz of observed bandwidth \citep{isaacson2017breakthrough}. 

We approximate a detected drift rate as linear on a waterfall plot\footnote{Rotational and orbital accelerations would typically be sinusoidal with time (not linear) but both can be approximated as linear so long as the observation is only occurring for a small portion of an orbit or rotation (order minutes). This assumption will be discussed further in Section \ref{sec: discussion}.} such that a search algorithm for linear features in image data is needed. Luckily, finding specific features in images is a common problem in image processing and analysis and such tools already exist.\footnote{These methods from image processing are rarely mentioned in a SETI context \citep[with a few exceptions, ][]{monari2006generalized, montebugnoli2006seti, fridman2011seti}.} For the purposes of this study, we focus on the Hough and Radon transforms \citep{hough1962method, radon1986determination} as well as a class of algorithms based on the tree summation method \citep{taylor1974sensitive,cullers1992seti, enriquez2017breakthrough} We discuss each of these methods below.

\subsection{A Summary of Current Methods \label{ssec: methods}}

\subsubsection{Hough and Radon Transforms}
\label{sssec: houghandradon}

A Classical Hough Transform can be used to search for linear features in an image. In the Hough Transform, each pixel in the image space is transformed into a sinusoid in a two-dimensional parameter space with axes of slope and intercept. After iterating through all of the pixels in the image, that parameter space will show maxima at the most likely slope-intercept values. A full prescription of this method can be found in eg. \citet{hough1962method, illingworth1987adaptive,van2004short}.

The advantages to the Classical Hough Transform are significant. There is no need for a priori information about the lines' position, duration, or slope, just the maximum slope chosen to limit the search. In addition, every point's contribution to the parameter space can be calculated independently, which makes this an ideal problem for the application of parallel processing. The process is robust to noise and occlusion (part of the feature being hidden from view) and finds all maxima simultaneously \citep{illingworth1987adaptive}. The algorithm is \(O(N^2 log N)\) complexity in image size and linear in drift rate \citep{vuillemin1994fast, gotz1995fast}.

However, there are some disadvantages to the Classical Hough Transform. The algorithm takes as inputs a range of slopes and intercepts and the resolution of each axis. For our purposes, the slope of a line in a waterfall plot is the drift rate, meaning that an upper bound for the drift rate is chosen by the user. This limits the search to linear features in a certain region of parameter space. Another disadvantage to this transform is its computational expense. Although there are some methods for improving the computational and memory requirements, the Hough Transform was for the most part ignored in image analysis as well as in SETI applications.

The Radon transform is a philosophical inversion of the Hough transform; they both have the same image space and parameter space. The Radon transform iterates through each pixel in the slope-intercept space instead of the image. Thus, the Radon transform is better for high resolution images, but much slower if you lack information about your parameters. This is explained in more depth by \citet{van2004short}. 

\subsubsection{Tree Summation Method}
\label{sss: dedoppler}

The tree summation method is more popular in SETI applications (eg. \citet{enriquez2017breakthrough}). This algorithm, originally used in the dedispersion of pulsars by \citet{taylor1974sensitive}, sums along all potential lines in an image up to a given slope. The summations for multiple drift rates involve redundant arithmetic, so the algorithm takes advantage of the redundancy by remembering parts of previous calculations. This makes the algorithm \(O(N log N)\) complexity in image size and linear in drift rate \citep{enriquez2017breakthrough}\footnote{Functionally, for discretized data, the linearity in drift rate is only true until the one-to-one point, discussed in Section \ref{ssec: discrete}, after which the speed improves. This caveat applies to both the Hough Transform and the tree summation method.}. Multiple wrappers have been developed for the tree summation algorithm to report the highest sum for a given frequency from a range of drift rates. \citet{enriquez2017breakthrough} uses \textit{turbo}SETI\footnote{\url{https://github.com/UCBerkeleySETI/turbo_seti}} for this task. A similar strategy, with the same complexities, is the Doubling Accumulation Drift Detector (DADD) used by \citet{cullers1992seti}.

\subsubsection{Coherent Methods}
\label{sssec: coherent}

All of the methods described above are 	``incoherent'': they are performed after the raw voltages are converted to a power spectrum, and thus only use the amplitude information of the signal. ``Coherent'' methods in the time-domain are also used --- the raw voltage data can be convolved with a chirp function before the conversion is performed, and thus it also uses the phase information of the wave. The coherent framework requires separate high resolution power spectra for each drift rate so the problem scales linearly with maximum drift rate. Unfortunately, coherent methods tend to be computationally infeasible for most searches. They are, however, used by the distributed computing pipeline of SETI@home \citep{anderson2002seti} and can be combined with incoherent methods to reduce the computational burden. Hybrid dedispersion methods for pulsar research already exist \citep{stappers2011observing} and could be extended to the problem of drift rates, but we will not explore them here. These methods coherently correct to a fiducial dispersion measure, then incoherently search around it.

Because the computation time for all of these methods scales positively with drift rate, picking a reasonable drift rate maximum directly affects the computation time required to perform the search, regardless of which search algorithm we use.\footnote{It should also be noted that all of the methods still have some degree of sensitivity to strong signals beyond their maximum drift rate cutoff; a piece of the signal can provide enough power to exceed a detection threshold.}

\subsection{The Beneficial Effects of Discrete Data \label{ssec: discrete}}

Real waterfall plots consist of pixels instead of a continuous space, providing some useful constraints on the range of drift rates that need to be searched. The format of the data itself provides hard limits on the minimum, maximum, and step size of the drift rate array. This has direct implications on the runtime of the search algorithm.

The minimum drift rate that can be distinguished is a signal that moves a total of one pixel (or bin) in frequency over the duration of the observation:

\begin{equation}
\label{eq: mindrift}
\dot{f}_{min} = \frac{\Delta f_{bin}}{t_{obs}}
\end{equation}

Any signal with a smaller drift rate than above will be indistinguishable from a signal with zero drift rate.

The maximum drift rate that can be distinguished is a signal that moves through only a single time bin over the bandwidth. Presented another way, this signal would be slewing so quickly in frequency space that it appears as a broadband flash. This limit is dependent on the time resolution and the total bandwidth of the receiver:

\begin{equation}
\label{eq: maxdrift}
\dot{f}_{max} = \frac{\Delta f_{bandwidth}}{\Delta t_{bin}}
\end{equation}

Multichannel receivers will capture all of the signal regardless of drift rate below this maximum, rendering much of the early SETI discussion about bandwidth choices for single-channel receivers inapplicable (see \citet{oliver1979rationale} for such a discussion).

Choosing the maximum drift rate is implicitly a problem of how the data are stored. By fixing the time and frequency resolution in a dynamic spectrum, the maximum drift rate that can be searched at a given sensitivity is also fixed. If the drift rate surpasses a one-to-one slope in time-frequency pixels ($\dot{f} \geq \frac{\Delta f_{bin}}{\Delta t_{bin}} $), the power will get smeared into multiple frequency bins in the same time bin, reducing the search's sensitivity to weak signals. Algorithms using the tree summation method can still be applied by manipulating the waterfall plots to bring the drift back below the new one-to-one point:  \citet{enriquez2017breakthrough} apply a linear frequency shift to the power spectrum as a linear function of time, while \citet{siemion20131} rebin the waterfall plot in the frequency dimension. The latter approach speeds up the search for drift rates past the one-to-one point because the resulting array size is smaller, but the sensitivity will still be a function of drift rate. If the raw voltage data are saved, a number of dynamic spectra that are coherently de-drifted to different (large) drift rates can be saved and incoherently searched to compensate for the sensitivity loss\footnote{This problem is closely related to hybrid de-dispersion methods in pulsar searches. This is illustrated by Figure 3 of \citet{Bassa2017}. Taken in a SETI context, the y-axis could be considered to be inversely proportional to sensitivity, the x-axis could be considered to be drift rate, and the number of minima in the semi-coherent line indicate the number of dynamic spectra that must be individually saved and searched.}. A complicated trade-off ensues between sensitivity, drift rate, computational storage, and computation time.

Finally, the drift rate step size, or maximum slope resolution that is useful to search, is just equal to the minimum drift rate in Equation \ref{eq: mindrift}. 

As an illustration, using the Breakthrough Listen setup on the GBT at 8 GHz \citep{macmahon2017recorder} produces a minimum drift rate and step size of about 0.009 Hz/s using the highest frequency resolution data. The maximum drift rate is about \( 2.4 \times 10^8\) Hz/s. This maximum drift rate is unreachable by a physical system that does not include a black hole or neutron star (see Section \ref{sssec: compact}); it creates an upper bound that is extremely high, and so requires extremely large amounts of computation time. The minimum drift rate and step size, on the other hand, are applicable and should be used as fundamental limits when constructing the parameter space for a Hough or Radon transform. The one-to-one point in this data is approximately 0.15 Hz/s, so the increase in computation speed by repeatedly halving the array size along the frequency axis greatly reduces the overall runtime \citep{enriquez2017breakthrough}. 

\section{Maximum Drift Rate Calculations for Representative Systems \label{sec: mdrexamples}}

\subsection{Solar System Bodies \label{ssec: solarsys}}

Using the equations derived in Section \ref{sec: math}, we searched the solar system for bodies that provide the largest drift rates and discuss the results below.

Although Equation \ref{eq: master} shows the full formulation of the drift rate, we normalize the drift rate by the rest frequency of the signal \(f_{rest}\) and report it in the unit of nanoHertz (nHz). Because this normalized quantity is independent of transmission rest frequency it frames the problem more intuitively as a drift in frequency as opposed to using pure accelerations and allows easy comparisons with previous work. For reference, the Earth's fractional drift rate from both rotation and orbital motion is 0.11 nHz. Other solar system bodies have similar drift rates: Mercury's orbital motion could impart a maximum drift rate of 0.13 nHz, Io's orbital motion around Jupiter could impart a maximum drift rate of 2.39 nHz, and Jupiter's rotational motion would impart a maximum drift rate of 7.2 nHz.

All of the values calculated in the following sections will be synthesized in Table \ref{tab: tabletoendalltables}, which could be consulted whenever a maximum drift rate needs to be chosen.

\subsubsection{2014 RC and 2008 DP4 \label{sssec: NEANEO}}

The Near Earth Asteroid 2014 RC is the fastest rotator in the solar system with a period of only 15.8 seconds \citep{nasa2014reports}. The object itself is extremely small, about 22 meters in equatorial extent \citep{nasa2014reports}. The drift rate from this object is 3.7 nHz.

Given that 2014 RC is so small, it might be expected that a slightly larger object rotating slightly slower could produce a higher drift rate. We searched the Asteroid Lightcurve Database \citep{warner2009asteroid} for all objects with a known period and diameter that had a rotation period of less than twelve hours. We calculated rotation-induced drift rates for all 13429 of these objects and found an object with an even faster drift rate. The outer main-belt asteroid 2008 DP4 (also known as 2003 HC33 and 2006 WW116) has a drift rate of 4.2 nHz from its 218.52 second period and diameter of 3.06 km \citep{warner2009asteroid}.

The period of 2014 RC is far shorter than the object's break-up period of 2.08 hours, as is 2008 DP4's break-up period of 1.91 hours. This indicates that both objects are not ``rubble piles'' held together by self-gravity, but rather monoliths held together by electrostatic forces in the rock itself.

\subsubsection{2006 HY51 \label{NEAellipse}}

Another way to get large accelerations is to look at highly-elliptical orbits such as those followed by minor bodies like comets and asteroids. Near periapse, the total acceleration will be maximized but objects spend only a small portion of their orbits in this position. The vis viva equation and Kepler's Third Law can be used to derive Equation \ref{periapse} for the fraction of an orbit spent near periapse, where \(\theta\) is a suitably small portion of the total orbit. With an eccentricity of \(e\) = 0.9 and a generous periapse angle of \(\theta = \frac{\pi}{2}\), the fraction of the orbit spent near periapse is 0.57\%. 

\begin{equation} \label{periapse}
\frac{\tau_{peri}}{P} = \frac{\theta}{2\pi}\frac{(1-e)^{3/2}}{(1+e)^{1/2}}
\end{equation}

Using data obtained from the IAU Minor Planet Center Orbit/Observation Database (MPC) \citep{center2018mpc}, we searched for objects with an eccentricity of $0.5 < e < 1.0$. A total of 8417 objects were returned. In order to maximize the observed accelerations, we used the periapse distance in Equation \ref{eq: general_orbit}. The MPC database includes uncertainty parameters to quantify their certainty of the values; we checked these for quality control. Following this calculation, we rejected the objects with the three highest drift rates on the basis of having high uncertainty parameters (9, 9, and 7, on a scale of 0-9, where anything above 6 is unusual) \citep{center2018mpc}. The fourth highest drift rate was from an object with an uncertainty parameter of 1: NEO 2006 HY51. It has a semimajor axis of 2.59 AU and an eccentricity of 0.97, creating a maximum orbital drift rate of 3.27 nHz.

\subsubsection{'Oumuamua}
\label{sssec: oumuamua}

The first interstellar asteroid within the solar system, 'Oumuamua, was discovered in 2017 \citep{Meech2017}. It is the only currently-known object in its class and thus provides a case-study for the example of a transmitter falling into the solar system. 'Oumuamua had a solar closest approach of 0.25 AU, which, by Equation \ref{eq: general_orbit} would give a maximum drift rate of 0.316 nHz. Such interstellar objects, just as with bound objects, would have to have extremely close approaches to the sun to produce large drift rates.

\subsection{Reasonable Extrapolations: Exoplanetary Systems \label{ssec: exoplanets}}

\subsubsection{Orbital Contributions: Small Semimajor Axes \label{sssec: semimajor}}

Using exoplanets.org \citep{han2014exoplanet} we obtained the 20 exoplanets with the smallest semimajor axes. From this list, we calculated a drift rate solely from the orbital motion of the planet. From a combination of the mass of the central star and the star-planet distance, the exoplanet with the largest circular orbital drift rate is Kepler-78b. At a distance of 0.0915 AU from its \(M = 0.83M_{sun}\) host star, it orbits with a period of 8.5 hours \citep{howard2013rocky}. This would cause a drift rate of 191 nHz. This is equivalent to 1531 Hz/s at 8 GHz. This planet is one of our closest Earth-analogues (20\% larger than Earth and 69\% more massive), with a measured density consistent with a rock-iron composition and a core size similar to Earth and Venus \citep{howard2013rocky}. The small semimajor axis makes it likely that the planet is tidally-locked, producing a negligible rotational component with an amplitude that is smaller by $\frac{R_P}{a}$. 

While the hot dayside temperature might not be amenable to the evolution of complex life, there are reasons that tidally locked planets with large associated drift rates such as Kepler-78b would be particularly interesting to SETI as good sites for ``beacons''. A larger fraction of the sky sees planets like this in transit because of the small semimajor axis. The short period allows them to cover the whole sky in a smaller amount of time. The planet itself would serve as a marker for the beacon's presence. Being close to the star would also provide a lot of available flux to power a transmitter on the night side of the planet, which could broadcast to the region of the sky seeing the planet transit at any given time \citep{kipping2016cloaking}.

\subsubsection{Orbital Contributions: Extremely Elliptical Orbits\label{sssec: elliptical}}

While circular orbits will produce periodic high drift rates for any sufficiently edge-on viewing angle, elliptical orbits have their drift rates maximized at a viewing angle aligned with the star and planet at periapse. As in the case of the comets in Section \ref{NEAellipse}, the drift rates were calculated for these viewing angles. Exoplanets.org \citep{han2014exoplanet} was searched for exoplanets with eccentricities greater than 0.7, returning 20 results. The resulting drift rates from this subset ranged from 0.05 nHz to 22.7 nHz. The 22.7 nHz drift rate came from HD 80606 b. This 3.9\(M_{jup}\) object has a semimajor axis of 9.4473 AU and a large eccentricity of 0.934, and it orbits with a 111.8 day period \citep{naef2001hd}. Although this object is a gas giant without a solid surface, such objects could host orbiting transmitters. It is also informative of maximum drift rates from exomoons around similar planets. In addition, a terrestrial planet on a similarly eccentric orbit is not physically impossible, though none have yet been observed.

\subsubsection{Rotational Contributions \label{sssec: rotational}}

Unfortunately, direct measurements of terrestrial exoplanet rotation rates will not be feasible until direct imaging instrumentation vastly improves in resolution. Rotation rates have been measured for dozens of brown dwarfs and less than 10 planetary-mass objects \citep{bryan2018constraints,scholz2018universal} (\(< 13 M_{Jup}\)). Of this handful of planetary mass objects with measured rotation rates, the maximum measured rotation rate is from \(\beta\) Pic B. \(\beta\) Pic B has a $log(g)$ surface gravity of 4.0, which, along with its \(M\) value, implies a radius of 1.5 \(R_{Jup}\) \citep{quanz2010first}. This gives a maximum rotational drift rate, given the radial velocity measurements of \cite{snellen2014fast}, of 19.4 nHz. This number might not have much real-world relevance, as a transmitter cannot be built on the surface of an 8.5\(M_{Jup}\) gas giant, but it is our only observational touchpoint for maximum rotational speeds attainable by exoplanets. In addition, it is the only exoplanet with a known rotation rate residing in a system with a debris disk. 

A side benefit of the detection of an ETI narrowband transmitter on a planetary surface would be the accurate measurement of planetary properties. This includes the planet's rotation period and a lower limit on the planet's radius, as well as a determination of the planet's orbital period, eccentricity, and other orbital parameters \citep{sullivan1978eavesdropping}\footnote{In fact, \citep{sullivan1978eavesdropping}, via examination of the Earth's radio signature, even raises the possibility of measuring the presence of a plasmasphere, the jitter from wind tilting the tallest transmitters, the inclination of the earth's axis, and the transmitters' antenna sizes.}.

Another way to constrain the rotational component of the drift rate would be to use a model of planetary formation to create a simulation that tracks the rotation rates of the simulated protoplanets as they evolve. \citet{miguel2010planet} found that the majority of planetary primordial rotation periods in their simulation fell between 10 and 10000 hours. An Earth-sized exoplanet with a rotation period of 10 hours would have a drift rate of 0.65 nHz, making this a softer upper limit than most derived from observed data in the previous section. Primordial spin rates can be affected by subsequent rotational evolution due to tidal processes or collisions \citep{hughes2003planetary}.

\subsection{Extreme Cases \label{ssec: extreme}}

Statements about exoplanet rotation rates and semimajor axes can be made based on observations of our own solar system and other exoplanetary systems. In order to be certain that all cases are considered, however, a physical hard upper limit is needed from theory. Here we consider extreme limiting cases for rotation, orbital motion around stars, and orbital motion around compact objects.

\subsubsection{Break-Up Speeds}
\label{sssec: breakup_calc}

For planetary rotation, the upper limit on drift rate is defined by the break-up rotation rate (derived in Section \ref{ssec: breakup}). We will use the definition of a super-Earth advanced by \citet{seager2007mass}; a super-Earth must be a solid planet larger than Earth with no significant gas envelope. There are many estimates of where the super-Earth cutoff would fall \citep{marcy2014masses, rogers2015most}; here we will use 6 Earth masses from \citet{dressing2015mass}). 

In accordance with standard practice in the exoplanet literature, we will now consider exoplanets with three different compositions for 6\(M_{\oplus}\) super-Earths: 100\% water ice, 100\% silicate (MgSiO$_3$ perovskite), and 100\% iron. Using corresponding radius values derived from the mass-radius relationships in \citet{Seager2007}, the drift rates from these three compositions are 44.4 nHz, 87.2 nHz, and 309 nHz, respectively.

These values are also the limiting drift rates for the maximum orbital contribution from an exomoon or orbiting transmitter bound to one of these bodies. In this case, the central body would not need to be solid, and we can maximize the drift rate around a gas giant by looking at the largest planet considered in \citet{Seager2007}: a 10\(R_{\oplus}\), 1300\(M_{\oplus}\) gas giant. This produces a maximum exomoon/transmitter orbital drift rate contribution of 424 nHz.

\subsubsection{Closest Allowable Orbits \label{sssec: closestorb}}

 The velocity of a transmitter on the equator of a sphere at break-up speed is identical to that of the maximum orbital velocity around the sphere. Thus, we can use Equation \ref{eq: Pbreakup} to calculate the upper limit on an exoplanet's orbital contribution to drift - its closest allowable orbit - based on the properties of its host star. In Table \ref{tab: stellar_table} we list the resulting drift rates from main sequence stars using values from \citet{zombeck2006handbook}. Note that, because masses and radii are known in this case, the exact density distributions can be disregarded.

\begin{table}[ht]
\centering
\begin{tabular}{|l|l|l|l|l|}
\hline
Stellar Type & Radius (\(R_{sun}\)) & Mass (\(M_{sun}\)) & Drift Rate (nHz) \\
\hline
O6           & 18                   & 40                 & 113              \\
B0           & 7.4                  & 18               & 301              \\
B5           & 3.8                  & 6.5                & 418              \\
A0           & 2.5                  & 3.2              & 468              \\
A5           & 1.7                  & 2.1             & 665              \\
F0           & 1.3                  & 1.7        & 920              \\
F5           & 1.2                  & 1.3            & 826              \\
G0           & 1.05                 & 1.1           & 913              \\
G2           & 1                    & 1                  & 915              \\
G5           & 0.93                 & 0.93              & 984              \\
K0           & 0.85                 & 0.78                   & 988              \\
K5           & 0.74                 & 0.69          & 1153             \\
M0           & 0.63                 & 0.47                 & 1083             \\
M5           & 0.32                 & 0.21                & 1876             \\
M8           & 0.13                 & 0.1              & 5413\\
\hline
\end{tabular}
\caption{Radius and mass information from \citep{zombeck2006handbook} and derived drift rates for an array of stellar types. The values are all quite extreme, representing the maximum observable drift rate from a transmitter orbiting at the radius of the host star. Mechanisms such as orbital decay and tidal disruption could prevent this limit from being reached in real physical systems, but are not considered here in the spirit of providing a general and absolute upper limit. With these values, drift rate maxima could be individually chosen for different targets in a survey based on the stellar type of the target. Consistent fractions of these maxima could also be chosen instead of the maxima themselves.}
\label{tab: stellar_table}
\end{table}

The values, especially for the later stellar types, are much larger than those calculated previously in this work. This indicates that the orbital motion drift rate contribution could, in a physically possible way, be much larger than that of known exoplanets.

Adding the physical limits of rotation of the host body, that body's orbital motion around a planet, and that system's orbital motion around a star yields a drift rate of 6146 nHz. However, this system is not physically realizable because it would correspond to a super-Earth-sized exomoon brushing the surface of a 1300\(M_{\oplus}\) gas giant, orbiting at the radius of its host M-dwarf star. It would also require a contrived orbital configuration relative to Earth and a very short and precisely timed observation window.

\subsubsection{Compact Objects\label{sssec: compact}} 

Black holes have their place in the SETI literature as (perhaps) a better-than-average location to search for highly technologically advanced intelligences because of the energy that could be extracted from them or the computation that could be accomplished near them \citep{penrose1969gravitational,vidal2011black}. It is not unthinkable that an ETI might be sending a beacon from a transmitter orbiting around a black hole \citep{Jackson}. Neutron stars and white dwarfs are discussed in the SETI literature as well \citep{cameron1963interstellar, Semiz2015, Osmanov2018, Imara} and are also considered below.

The methods used in this section are purely Newtonian, but a general relativistic approach will yield values of the same orders of magnitude for the specific problems considered below. Looking at the masses for both stellar \citep[e.g., Cygnus X-1, \(14.8 M_{\odot}\)][]{boehle2016improved} and supermassive \citep[e.g., Sagittarius A*, \(4.02 \times 10^6 M_{\odot}\)][]{orosz2011mass} black holes, we can calculate the drift rates from transmitters at the innermost stable circular orbit (ISCO)\footnote{Assuming non-rotating black holes}. For white dwarfs and neutron stars, we use order of magnitude values (\(M_{wd} = 0.6 M_{sun}\), \(M_{ns} = 1.4 M_{sun}\), \(r_{wd} = r_{earth}\), \(r_{ns} = 10\) km)  and set the closest allowable orbit as the maximum.

The drift rates are predictably enormous. For Sagittarius A*, the drift rate is \(4.7 \times 10^5\) nHz. The typical white dwarf values produced \(6.5 \times 10^6\) nHz. For Cygnus X-1, the drift rate is \(1.3 \times 10^{11}\) nHz. The typical neutron star values produced \(6.2 \times 10^{12}\) nHz. The signal from Cygnus X-1 and the neutron star drift so quickly that they would show up as a broadband pulse in Breakthrough Listen data (see Section \ref{ssec: discrete}), and a weak one at that (resulting from how quickly they would pass through the band compared to the integration time). This implies that potential signals from compact objects would need specific consideration to be detectable with current methods.

\begin{table}[ht]
\centering
\begin{tabular}{|l|l|l|l|}
\hline
Situation & Object & Fractional Drift Rate (nHz) & Section \\
\hline
Solar System - Terrestrial Planet - Earth's Contribution                                & Earth       & 0.11        & \ref{ssec: solarsys}               \\
Solar System - Terrestrial Planet - Observed                                & Mercury       & 0.13        & \ref{ssec: solarsys}               \\
Solar System - Interstellar Asteroid - Observed & 'Oumuamua & 0.136 & \ref{sssec: oumuamua}\\
Simulation - Terrestrial Planet - Common Fast Rotator & \nodata & 0.65 & \ref{sssec: rotational}\\
Recommended Value - \citet{oliver1971project} & \nodata & 1.0 & \ref{sec:intro} \\
Solar System - Moon - Observed                                              & Io            & 2.39  & \ref{ssec: solarsys}                      \\
Solar System - NEO (Highly Eccentric) - Observed                            & 2006 HY51     & 3.27   & \ref{NEAellipse}            \\
Solar System - Asteroid (Fast Rotator) - Observed                           & 2008 DP4      & 4.22  & \ref{sssec: NEANEO}           \\
Solar System - Gaseous Planet - Observed                                    & Jupiter       & 7.2   & \ref{ssec: solarsys}                      \\
Exoplanet - Rotational - Observed                                           & \(\beta\) Pictoris b & 19.4  &  \ref{sssec: rotational} \\
Exoplanet - Highly Eccentric - Observed                                     & HD 80606b     & 22.7         &   \ref{sssec: elliptical}  \\
Exoplanet - Rotational - Terrestrial Upper Limit (H\(_2\)O)                               & \nodata             & 44.4   & \ref{sssec: breakup_calc} \\
Exoplanet - Rotational - Terrestrial Upper Limit (MgSiO$_3$)                             & \nodata            & 87.2 & \ref{sssec: breakup_calc}  \\
Exoplanet - Small Semi-Major Axis - Observed                                & Kepler-78 b   & 191  & \ref{sssec: semimajor}      \\
\textbf{Recommended Value - This Work} & \nodata & \textbf{200} & \ref{sec: conclusion}\\
Exoplanet - Rotational - Terrestrial Upper Limit (Fe)                               & \nodata             & 309   & \ref{sssec: breakup_calc} \\
Exoplanet - Rotational - Gaseous Upper Limit (H/He)                               & \nodata             & 424   & \ref{sssec: breakup_calc} \\
Exoplanet - Orbital - G2 Stellar Upper Limit                           & \nodata             & 915  & \ref{sssec: closestorb}     \\
Exoplanet - Orbital - M8 Stellar Upper Limit                            & \nodata            & 5413   & \ref{sssec: closestorb}   \\
System - Exoplanet + Exomoon + Rotation - Upper Limit & \nodata             & 6146 & \ref{sssec: closestorb} \\
Supermassive Black Hole - Orbital - ISCO Upper Limit & Sagittarius A* &	\(4.7 \times 10^5\) & \ref{sssec: compact} \\
White Dwarf - Orbital - Upper Limit & \nodata & \(6.5 \times 10^6\) & \ref{sssec: compact}\\
Stellar Mass Black Hole - Orbital - Upper Limit & Cygnus X-1 & \(1.3 \times 10^{11}\) & \ref{sssec: compact}\\
Neutron Star - Orbital - Upper Limit & \nodata & \(6.2 \times 10^{12}\) & \ref{sssec: compact}\\
\hline
\end{tabular}
\caption{A summary of the results of this paper, shown in tabular form. Each row contains and describes a specific physical system, gives the object from which the parameters were taken (if applicable), gives the associated drift rate, and provides a pointer to the section where the system is discussed in detail. Rows are sorted by increasing drift rate --- an example visualization of some of the upper rows is shown in Figure \ref{img: onlyfigure}. When a maximum drift rate is chosen for a study, it can be compared with this table; all situations corresponding to rows above the chosen drift rate would be captured in the search while all rows below would be outside of the scope of the search. ``Simulation'' is based on \citep{miguel2010planet} as described in Section \ref{sssec: rotational}. ``Upper Limit'' rows are based on the cases described in Section \ref{ssec: extreme}. Note three things: 1) That the drift rates that could be produced by known astrophysical systems greatly exceed the maximum drift rates chosen by many SETI searches in the past 2) Given the better-than-linear scaling in drift rate past the one-to-one point (see Section \ref{ssec: discrete}), using a more physically-motivated maximum drift rate is computationally feasible 3) Some amount of subjectivity must still go into the choice of a maximum drift rate, but the authors recommend \textbf{200 nHz} to account for the possible drift rates produced by all observed solar system bodies and exoplanets.}
\label{tab: tabletoendalltables}
\end{table}

\begin{figure}[ht]
\centering
\includegraphics[width=0.8\textwidth]{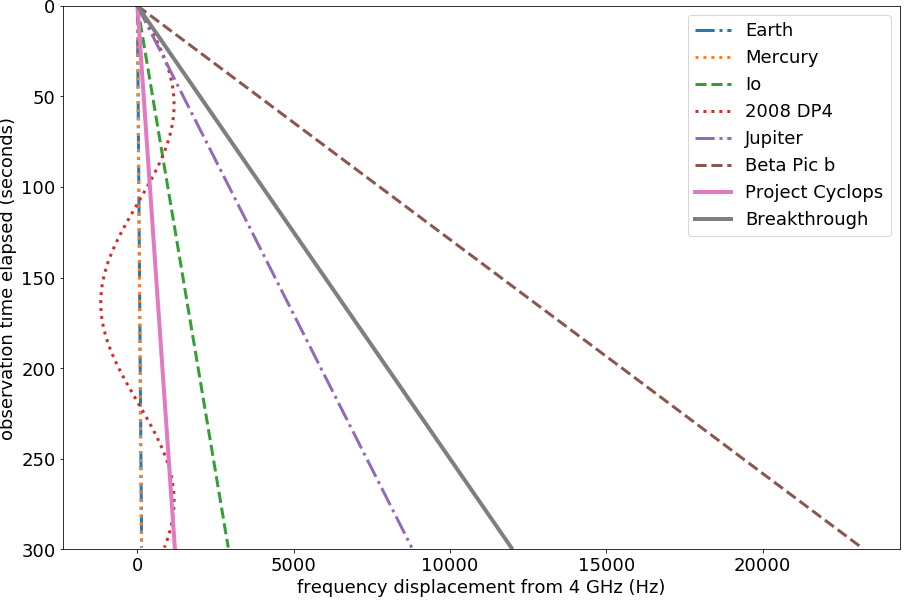}
\caption{A simulation of a five minute observation of narrowband signals with various drift rates. The signal in the frame of the transmitter is a constant, narrowband 4 GHz transmission. The difference in profiles is due solely to the relative acceleration between the transmitter and the receiver. At t=0, all signals are at the maximum drift rate (minimum radial velocity and curvature) portion of their motion. Dotted, dashed, and dot-dashed lines indicate various bodies from from Table \ref{tab: tabletoendalltables}. The two solid lines indicate the highest drift rates considered by two SETI programs (those used in \citet{oliver1971project} and \citet{enriquez2017breakthrough}). Maximum drift rates produced by the signals to the right of the solid lines (counter-clockwise rotation) would not be detected by those SETI searches. Earth and Mercury overlap as near-vertical lines in this illustration. The sinusoidal motion of 2008 DP4 is visible due to its short period.} 
\label{img: onlyfigure}
\end{figure}

\section{Discussion \label{sec: discussion}}

One benefit of drift rate analysis is that it provides the searcher with built in rejection for radio frequency interference (RFI). Any signal with precisely zero drift rate is not accelerating radially with respect to the receiver, and so is likely also on the Earth's surface. This drift rate based RFI rejection has been performed in SETI searches using the Allen Telescope Array as well as the Breakthrough Listen Project \citep{harp2016seti, sheikhinprep}. SETI Institute searches such as \citet{harp2016seti} not only flagged zero drift rate signals, but also those with ``drift rate too high'' as RFI coming from a passing satellite. 

One consequence of the galactic drift rate contribution being negligible is that if a transmitter in the Milky Way corrects for the rotational and orbital motion of their own system, the only drift rate imparted on the signal will be from the rotational and orbital motion of Earth. This idea, that an ETI will account for their own drift rate in order to put their signal in the drift rate Schelling Point \citep{wright2017exoplanets} of the galactic barycenter, has been around for a while \citep{drake1984seti,horowitz1993five,leigh1998interference}. A caveat to this idea is that the signal can only be de-drifted for a single direction at a time; even if the center of the transmission beam is de-drifted, the edges of the beam will still show a drift rate. It would be difficult to build an isotropic transmitter that does not produce a drift in any direction. However, with the placement of a stationary beacon far outside the gravitational influence of any star system (i.e. strategically manipulated to have zero acceleration in the galactic barycentric frame) one could produce a de-drifted isotropic beacon.

Transmitters might serve functions other than as beacons to humanity and might naturally be found in high-acceleration environments because of their utility for energy generation or computation (see Section \ref{sssec: compact}). Transmitters could also have additional sources of frequency drift due to rotational accelerations from the transmitter itself. Much like the cases of 2014 RC and 2008 DP4 in Section \ref{sssec: NEANEO}, small enough objects are not subject to the break-up speed as a limit and upper limits due to material strength would be difficult to place.

The absolute value of the drift rate informs us about the acceleration of the system in question. A signal that exhibited an unphysically large drift rate might indicate a narrowband signal that is purposefully being swept in frequency, giving us a way to distinguish it from typical astrophysical signals \citep{fridman2011seti}. Unfortunately, this idea by its very nature would force a higher drift rate limit than discussed in this paper, with no obvious upper bound. While it is usually unwise to assume anything about the motives of an ETI, this sort of signal would slip through the algorithmic nets that we use today and would seem, therefore, to be a poor way for an ETI to intentionally broadcast its presence.

Even though, mathematically, positive and negative drift rates are symmetrical, physically, another constraint arises: radio telescopes can only observe above the horizon. Consequently, most drifts from Earth's rotation should be negative (drifting to lower frequencies over time). This means that signals with positive drift rates might warrant a closer look than those with negative drift rates.

One point that was not discussed elsewhere in this work is that non-controlled oscillators have a frequency drift that can be quite large. Terrestrial sources of interference that are poorly temperature controlled, typically those with low-cost components, will therefore sometimes appear as drifting signals even though they are in the same inertial frame as the receiver.

This work focused on narrowband detection algorithms, but linear drift rates still apply to arbitrary signal modulations such as broadband signals or combs, and these modulations would be missed by the methods in Section \ref{sec: computation}. Cross-correlation functions, however, would be sensitive to a much wider range of modulations. Using the cross-correlation function in a SETI project would involve an initial RFI masking for anything with no drift rate, and then running the cross-correlation between adjacent pairs of spectra and looking for peaks; this has been performed in \citet{harp2015radio}. In addition, convolutional neural networks (CNNs) have recently been used to identify pulses in the repeating Fast Radio Burst FRB121102 \citep{zhang2018fast} and in anomaly detection for SETI \citep{zhang2019self, harp2019machine}. Giving a CNN a training set of radio frequency interference (RFI) and empty frequency channels would allow it to classify data of these types, and a flag would be raised whenever it encountered an ``anomaly''. This is an appealing technique because it allows all drift rates and all signal types to be searched - the algorithm is just looking for something ``anomalous''. However, anomaly detection is still a nascent sub-field of machine learning and much work needs to be done to develop and test these algorithms.

Finally, it is noted in many SETI papers \citep[e.g.,][]{monari2006generalized, siemion20131, enriquez2017breakthrough} that the drifting signal, assuming that the drift is caused by periodic rotational or orbital motion, would actually appear sinusoidal if observed for long enough periods of time. In shorter, order minutes, observations, the non-linearity can be neglected. However, longer observations looking for a square root gain in sensitivity with time might eventually need to account for this sinusoidal behavior. If similar techniques were applied to this case, the fitting of the sinusoid would require a Hough Transform into a four-dimensional parameter space \citep{monari2006generalized}, with two of those dimensions (period and amplitude) associated with the maximum accelerations of the physical system of the transmitter. Thus, the choice of ``maxima'' and the search for more efficient algorithms are even more applicable to this more complex case.

\section{Conclusion \label{sec: conclusion}}

Though the physics behind the existence of a drift rate in a narrowband signal are quite simply derived from the classical Doppler shift equation, the application of the physics to the way drift rates are searched in SETI has been lacking a firm foundation. We have here provided that foundation.

All commonly used algorithms for finding a drifting signal (Hough Transforms, Radon Transforms, and the tree summation technique) scale (for constant SNR) linearly with drift rate until the one-to-one point (Section \ref{ssec: discrete}) and better than linearly beyond that point (though in a trade off with SNR). The resulting unavoidable connection between the computational resources required for a search and the range of drift rates searched drive us to consider which drift rates are too high to be worth searching for.

In this paper, using the formulae derived in Section \ref{sec: math}, we examined the drift rates that would be produced based on objects in our solar system, from planets to moons to asteroids and comets. We then applied similar techniques to exoplanets with known semimajor axes, eccentricities, and, to the amount possible, rotations. Finally, we looked at extreme cases: the break-up rotation speeds of plausible extrasolar objects, the closest allowable orbits of extrasolar systems, and a few exotic cases to illustrate the way that we could maximize radial accelerations with known objects.

Based on Table \ref{tab: tabletoendalltables}, a few conclusions can be drawn. 

\begin{itemize}

\item The drift rates that could be produced by known astrophysical systems greatly exceed the maximum drift rates chosen by many SETI searches in the past. \citet{oliver1971project} proposed a fractional rate of 1 nHz. Comparison with Table \ref{tab: tabletoendalltables} indicates that this search would have missed signals from Io and any subsequent rows.

\item Given the initially linear scaling in drift rate, using a more physically-motivated maximum drift rate will be more computationally intensive. However, the improvement in search speed past the one-to-one point (see Section \ref{sss: dedoppler}) will cause the use of the 200 nHz guideline to be within the computational capabilities of current large radio SETI searches, if not yet in real-time analysis.

\item Some amount of subjectivity must still go into the balance between computational expense and search completeness. It is still up to the author of a future SETI survey to decide which systems are so implausible as to no longer warrant the additional computational time. Said another way, there is no objective way to decide which row of Table \ref{tab: tabletoendalltables} is the ``cut-off row'', given our current knowledge of exoplanetary systems, our guesses about the habits of an ETI, and the specific goals and targets of a hypothetical search. That said, we recommend \textbf{200 nHz} as a value that would encompass the possible drift rates produced by all observed solar system bodies and exoplanets.

\item The choice of a maximum drift rate is not unique to radio SETI; An analogous problem appears in choosing a maximum dispersion measure in pulsar searches. Signal detection algorithms in radio SETI have much in common in a wide range of image processing applications. Though much of this paper is specific to SETI, the consideration of radial accelerations and the algorithmic search for signals have many applications throughout astrophysics and aerospace engineering (eg. spacecraft communications).

\end{itemize}

\section{Acknowledgements \label{sec: acknowledgements}}

This research was partially supported by Breakthrough Listen, part of the Breakthrough Initiatives sponsored by the Breakthrough Prize Foundation\footnote{\url{http://www.breakthroughinitiatives.org}}.
The Center for Exoplanets and Habitable Worlds is supported by the
 Pennsylvania State University, the Eberly College of Science, and the
 Pennsylvania Space Grant Consortium. The authors would like to acknowledge Eric Ford and Robert Collins for enlightening conversations and contribution to Section \ref{sec: computation}, as well as Andrew Shannon for contribution to Section \ref{sssec: rotational}. The authors would also like to acknowledge Jill Tarter for her contributions throughout this work and Mariah Macdonald for contributions during the editing phase. This research has made use of data and/or services provided by the International Astronomical Union's Minor Planet Center. SZS thanks the SETI Institute for providing a workspace for summer 2018, and Breakthrough Listen for providing the funding for this work. SZS acknowledges Neil Peart for the choice of Cygnus X-1.

\bibliography{references,more_references}
\bibliographystyle{aasjournal}

\end{document}